\documentclass[fleqn,usenatbib]{mnras}

\usepackage{newtxtext,newtxmath}

\usepackage[T1]{fontenc}

\DeclareRobustCommand{\VAN}[3]{#2}
\let\VANthebibliography\thebibliography
\def\thebibliography{\DeclareRobustCommand{\VAN}[3]{##3}\VANthebibliography}

\usepackage{graphicx}	
\usepackage{amsmath}	
\usepackage{subcaption}
\usepackage{xcolor}

\title[Project Hephaistos I.
Upper limits on partial Dyson spheres]{Project Hephaistos I. Upper limits on partial Dyson spheres in the Milky Way}

\author[Matías Suazo et al.]{
Matías Suazo,$^{1}$\thanks{E-mail: matias.suazo@physics.uu.se}
Erik Zackrisson,$^{1}$
Jason T. Wright,$^{2,3,4}$
Andreas J. Korn,$^{1}$
Macy Huston$^{2,3,4}$
\\
$^{1}$Observational Astrophysics, Department of Physics and Astronomy, Uppsala University, Box 516, SE-751 20 Uppsala, Sweden\\
$^{2}$Department of Astronomy \& Astrophysics, The Pennsylvania State University, University Park, PA 16802\\
$^{3}$Penn State Extraterrestrial Intelligence Center, The Pennsylvania State University, University Park, PA 16802\\
$^{4}$Center for Exoplanets and Habitable Worlds, The Pennsylvania State University, University Park, PA 16802 
}

\date{Accepted 2022 January 22. Received 2022 January 21; in original form 2021 October 22
}

\pubyear{2015}

\begin{document}
\label{firstpage}
\pagerange{\pageref{firstpage}--\pageref{lastpage}}
\maketitle

\begin{abstract}
Dyson spheres are hypothetical megastructures built by advanced extraterrestrial civilizations to harvest radiation energy from stars. Here, we combine optical data from Gaia DR2 with mid-infrared data from AllWISE to set the strongest upper limits to date on the prevalence of partial Dyson spheres within the Milky Way, based on their expected waste-heat signatures. Conservative upper limits are presented on the fraction of stars at $G\leq 21$ that may potentially host non-reflective Dyson spheres that absorb 1--90\% of the bolometric luminosity of their host stars and emit thermal waste-heat in the 100--1000 K range. Based on a sample of $\approx 2.7\times 10^5$ stars within 100 pc, we find that  a fraction less than $\approx 2\times 10^{-5}$ could potentially host $\sim 300$ K Dyson spheres at 90\% completion. These limits become progressively weaker for less complete Dyson spheres due to increased confusion with naturally occurring sources of strong mid-infrared radiation, and also at larger distances, due to the detection limits of WISE. For the $\sim 2.9\times10^8$ stars within 5 kpc in our Milky Way sample, the corresponding upper limit on the fraction of stars that could potentially be $\sim 300$ K Dyson spheres at 90\% completion is $\lesssim 8\times 10^{-4}$.
 
\end{abstract}




\section{Introduction}
The construction of swarms of artificial, light-absorbing components around stars could in princple allow advanced extraterrestrial civilizations to harvest vast amounts of radiation energy in the form of star light \citep{dyson60}. The  observational signatures of such ``Dyson spheres'' include waste-heat from the absorbing sphere, obscured direct star light (resulting in both an apparently underluminous star and potential temporal variations in brightness) and the effects of  feedback on the properties of the star from the surrounding sphere \citep[e.g.,][]{dyson60,wright16,Wright20}. 

Unlike searches for technologically advanced civilizations based on communication signals, searches based on Dysonian signatures do not strongly hinge on the willingness of other civilizations to contact us. On the nine axes of merits for technosignature searches defined by \citet{Sheikh20}, waste-heat signatures rate favourably on detectability and cost (because searches can largely be done with existing data), ancillary benefits (because candidates that turn out not to be Dyson spheres may still be interesting for mainstream astrophysics), inevitabitily (because the emission of thermal waste-heat radiation is difficult to evade from thermodynamic considerations) and duration (Dyson spheres can be very long-lived). 

A number of observational searches for individual Dyson spheres \citep[e.g.,][]{Slysh85,Jugaku91,Timofeev00,Jugaku04,Carrigan09,Zackrisson18} and for the large-scale use of Dysonian technology at extragalactic distances have previsously been carried out \citep{Annis99,Wright14a,Wright14b,Griffith15,Zackrisson15,Garrett15,Lacki16,Olson17,Chen21}, although without uncovering any truly compelling Dyson sphere candidates. In terms of sample size, the largest searches for individual Dyson spheres \citep[e.g.,][]{Slysh85,Jugaku91,Timofeev00,Carrigan09} are based on the IRAS catalog \citep{IRAS}, which includes $\approx 2.5\times 10^5$ point sources. However, modern photometric surveys covering optical, near-infrared and mid-infrared wavelengths reach object counts of up to $\sim 10^9$ targets, and should allow for even larger search programmes. 

As part of {\it Project Hephaistos\footnote{\url{https://www.astro.uu.se/~ez/hephaistos/hephaistos.html}}}, we intend to carry out the largest search to date for partial Dyson spheres, using data on $\sim 10^8$ Milky Way objects from combined Gaia, 2MASS and WISE catalogs. While Gaia, 2MASS and WISE all provide photometric data, Gaia crucially also provides parallax-based distances, which allow the spectral energy distributions of the targets to be anchored to an absolute luminosity scale.

In this first paper, we explore the upper limits that can be set on the prevalence of near-complete (capturing $\gtrsim 10\%$ of the radiation from their host stars) 100-1000 K Dyson spheres at distances up to $\approx 5$ kpc based on their expected waste-heat signatures, using a combination of Gaia and WISE data. In a later paper, we will take a closer look at the targets with the most Dyson sphere-like spectral energy distributions.

While it has been proposed that Dyson spheres and similar radiation-harvesting megastructures could be constructed around a variety of stellar-mass objects, including white dwarfs \citep{Semiz15}, pulsars \citep{Osmanov16,Osmanov18} and black holes \citep{Hsiao21}, we here limit the discussion to Dyson spheres around main sequence stars. We furthermore assume the Dyson spheres to be non-reflective and that feedback from the Dyson sphere on the star \citep{Huston21} may be neglected.

Our overall method is described in Section~\ref{sec:methods} and the resulting Dyson sphere limits in Section~\ref{sec:results}. In Section~\ref{sec:discussion} we discuss the likely nature of some of the Dyson-sphere interlopers uncovered in this work and also explore the impact that some of the underlying assumptions in our Dyson sphere models may have on our upper limits. Section~\ref{sec:conclusions} summarizes our results.

\section{Methods}
\label{sec:methods}

In this paper, we combine data from Gaia Data Release 2 \citep{gaia_space,gaia_dr2} and AllWISE \citep{cutri}, where Gaia DR2 provides parallaxes and fluxes in three optical bands ($G_\mathrm{BP}$, $G$, $G_\mathrm{RP}$) and AllWISE mid-infrared (mid-IR) fluxes at 3.4, 4.6, 12, and 24 $\mu m$. The AllWISE program is an extension of the WISE program \citep{wright_wise} and combines data from different phases of the mission.

To place upper limits on fraction of main-sequence stars that could potentially host Dyson spheres (DS), we define a set of colour-magnitude diagram (CMD) regions where objects compatible with DS are expected to turn up, and count the number of candidates falling into these regions for different distance-based subsamples. In Section~\ref{sec:models} we describe the models used to predict the positions of DS in these CMDs. In Sections~\ref{sec:spectrum} and~\ref{sec:photometry} we show some examples of how our models modify the spectrum and the photometric broadband fluxes of stars, and how the choice of DS model parameters affect the output. In Section~\ref{sec:counting_method}, we explain the counting method used to impose our upper limits. Finally, in Section~\ref{sec:data_samples} we demonstrate how the models work on a sample of stars and explain some observed trends.




\subsection{Dyson sphere models}
\label{sec:models}
When predicting the observational signatures of the composite system formed by a star and its DS, we model the stellar component as an obscured version of its original spectrum and the DS component as a blackbody whose brightness depends on the amount of radiation it collects.


The AGENT formalism described in \cite{Wright14b} represents a useful way of parameterizing the different energy inputs and outputs of DS. Here, AGENT is mnemonic for $\alpha \gamma \epsilon \nu T_{\text{waste}}$, where $\alpha$ represents the energy collected by the DS, $\epsilon$ the energy collected by other means (e.g., energy resources from their planet), $\gamma$ the thermal waste-heat (for a given characteristic temperature $T_{\text{waste}}$) and $\nu$ other losses (e.g., neutrinos, gravitational waves, etc.). Energy conservation is described by Equation~\ref{eq:agent} under this parameterisation:


\begin{equation}
\centering
    \alpha + \epsilon = \gamma + \nu.
    \label{eq:agent}
\end{equation}

Throughout this paper, we assume that the civilization's main energy input comes from the stellar radiation captured by the DS ($\epsilon \ll \alpha$) and that the main form of energy disposal is through thermal photons ($\nu \ll \gamma$). Under these assumptions, the energy conservation and steady-state  Equation~\ref{eq:agent} becomes Equation~\ref{eq:assumption}:


\begin{equation}
    \alpha = \gamma.
    \label{eq:assumption}
\end{equation}

Additionally, we assume that DS act like gray absorbers and that the thermal loss resembles a blackbody with a temperature in the range of 100 - 1000 K.  This range of temperatures is set by the detectability of the infrared excess radiation from the DS in the mid-IR wavelength range where WISE operates. Temperatures that are lower than 100 K would place this excess in the far-IR, whereas temperatures above 1000 K fall would shift it into the near-IR. 

Under these assumptions, only two parameters are required to model the spectra of stars surrounded by DS: the effective temperature $T_{\text{DS}}$ of the DS and the luminosity of the thermal radiation released by the DS. For the sake of simplicity, we re-define $\gamma$ as the normalized DS energy output: 
\begin{equation}
\gamma = \frac{L_{\rm DS}}{L_{\star}},    
\end{equation} where $L_{\rm DS}$ is the luminosity of the DS and $L_{\star}$ is the luminosity of the star hosting the DS before being obscured. Under this definition, $\gamma$ can take values between 0 and 1. In the case of an isotropically radiating star, $\gamma$ also represent the fractional solid angle of outgoing radiation intercepted by the DS (the covering factor) or, in more casual terms, the level of completion of the DS. Because of this, we use $\gamma$ and covering factor interchangeably throughout this paper.


\subsection{Spectrum}
\label{sec:spectrum}
With the assumptions above, we can model the photometric and spectroscopic properties of an obscured star and its DS. In this section, we present a basic spectral model for this compound system.


The overall spectrum of this system corresponds to the sum of the spectral contribution from each component. The intrinsic shape of the star's spectrum is unchanged, but its specific luminosity dims by a factor (1 - $\gamma$) due to the gray absorber assumption. Since the DS spectrum is considered a blackbody at an assumed effective temperature, the specific luminosity of the combined system becomes:


\begin{equation}
L_{\nu} = (1 - \gamma)L_{\nu,\star} + \gamma BB_{\nu}(T_{\rm DS},L_{\rm DS}),  \label{eq:spectrum}
\end{equation} where $L_{\nu}$ is the specific luminosity of the combined DS+star system, $L_{\nu,\star}$ is the specific luminosity of the star before being obscured, $\gamma$ is the covering factor, $L_{\rm DS}$ is the bolometric luminosity of the DS, $T_{\rm DS}$ is the temperature of the DS, and $BB_{\nu}(T,L)$ is the specific luminosity for a blackbody-like source with a given temperature ($T$) and bolometric luminosity ($L$).

\begin{figure}
    \centering
    \begin{subfigure}[b]{0.48\textwidth}
    \centering
    \includegraphics[width=\textwidth]{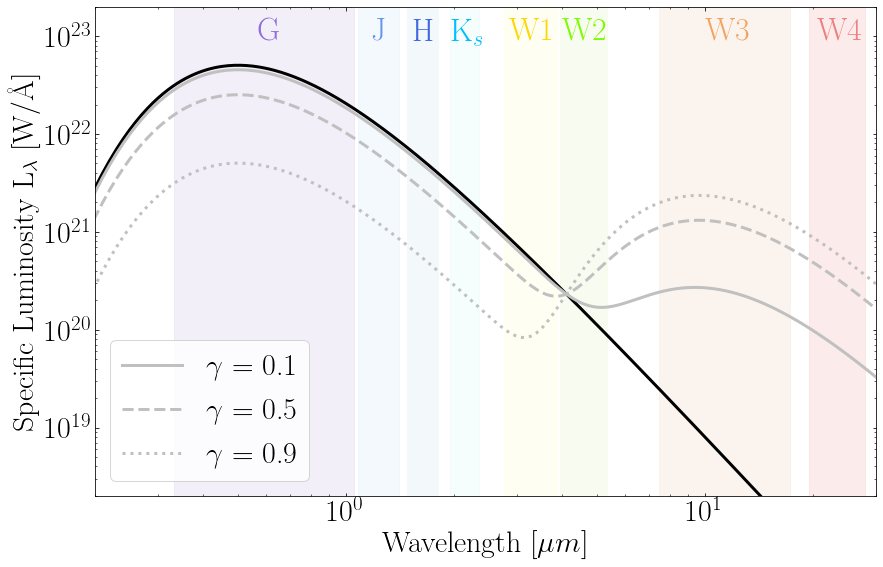}
    \end{subfigure}
    \begin{subfigure}[b]{0.48\textwidth}
    \centering
    \includegraphics[width=\textwidth]{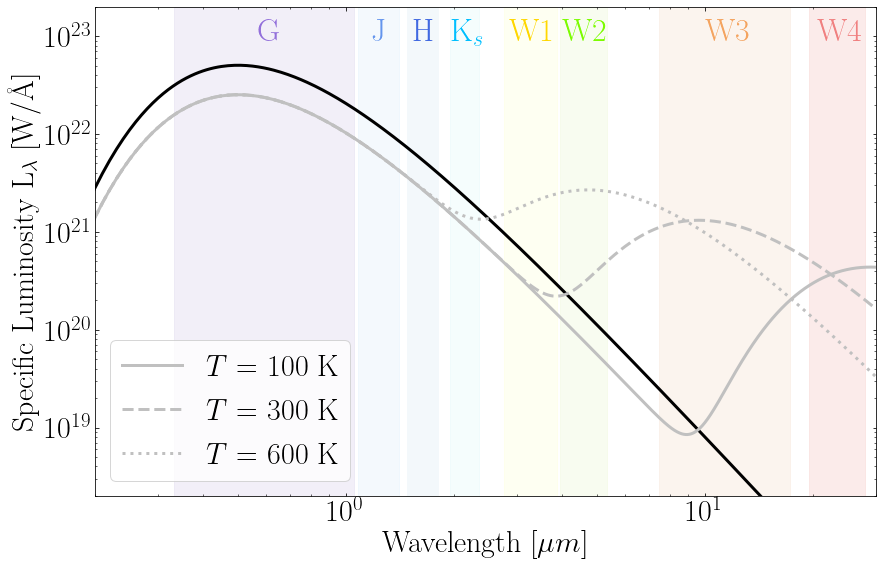}
    \end{subfigure}
     \caption{Modified Sun-like blackbody spectra due to the presence of different Dyson spheres. In both panels, the black solid line represents the unmodified blackbody ($T_{\text{eff}}$ = 5778 K, $L_{\star}$ = 1 L$_{\odot}$). In the top panel, DS models with a temperature of $T = 300$ K and covering factors of 0.1, 0.5, and 0.9 are depicted with grey solid, dashed, and dotted lines respectively. In the bottom panel, DS models with a covering factor of 0.5 and temperatures of 100, 300, and 600 K are depicted with grey dotted, dashed, and solid lines, respectively. The coloured bands represent the wavelength intervals relevant for the Gaia, 2MASS, and WISE missions.}
    \label{fig:spectra}
\end{figure}

Figure~\ref{fig:spectra} presents some illustrative examples of how a Sun-like blackbody spectrum ($T_{\text{eff}} = 5778$ K and $L_{\star} = 1$ L$_{\odot}$) is modified in the presence of a DS, under different assumptions about the DS temperature and covering factor. In both panels, a Sun-like blackbody spectrum is shown. In the top panel, we demonstrate how the compound spectrum changes due to a DS with a fixed temperature of 300 K and a covering factor of either $\gamma = 0.1$, 0.5, and 0.9. As seen, the DS causes a boost in the mid-infrared and a drop in the luminosity of the stellar component. Both features depend on the DS covering factor. The drop is predominant in the optical and the near-infrared. In the bottom panel, we show how the spectrum varies when we apply DS models with a fixed covering factor of 0.5 and DS temperatures of 100, 300, and 600 K. The signatures previously mentioned are recovered: a drop in stellar flux and a boost in the mid-infrared. Since we are considering various DS temperatures, the mid-infrared blackbody peak shifts in wavelength depending on the temperature of the DS. However, for the temperature range chosen here, the excess remains in the detectable wavelength window of the WISE mission.


\subsection{Photometry}
\label{sec:photometry}


Given our previous assumptions, equation~\ref{eq:mag} represents the magnitude ($M$) of the combined star+DS system. This magnitude depends on the DS's magnitude ($M_{\rm DS}$) and the star's magnitude after being obscured ($M_{\star}$) by the DS. This equation can be generalized to any photometric band,  and applies to both apparent and absolute magnitude. For the rest of this section, we use absolute magnitudes. Throughout the rest of the paper, all magnitudes are given in the Vega system.


\begin{equation}
      M = -2.5\log(10^{-M_{\star}/2.5} + 10^{-M_{\rm DS}/2.5})
      \label{eq:mag}
\end{equation}

Here, $M_{\rm DS}$ represents the absolute magnitude of the DS that is calculated theoretically by integrating the blackbody spectrum. The spectrum of this blackbody depends on the temperature ($T_{\rm DS}$) and the bolometric luminosity ($L_{\rm DS} = \gamma L_{\star}$). Here, $T_{\rm DS}$ and $\gamma$ are considered free parameters. However, the bolometric luminosity of the star is required as an input.

On the other hand, the magnitude of the combined system requires deriving the star's brightness after being obscured by the DS structure. Under the assumption of a gray absorber, the obscured magnitude of the star becomes \begin{equation}
      M_{\star} = M_{\star,O} -2.5\log(1 - \gamma)
      \label{eq:mstar}, 
\end{equation} where $M_{\star}$ is the magnitude of the star after being obscured, $M_{\star,O}$ is the magnitude of the star before being obscured, and $\gamma$ is the covering factor.

\begin{figure}
	\includegraphics[width=\columnwidth]{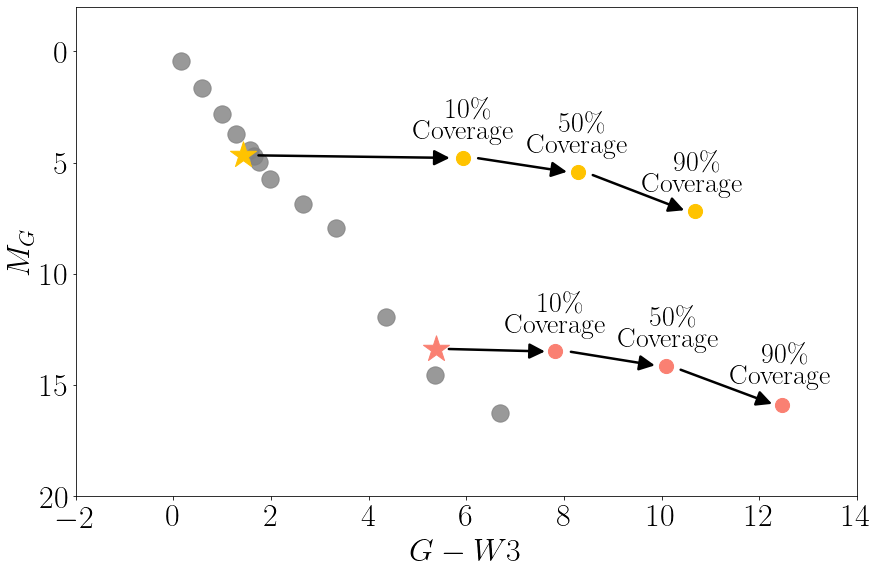}
    \caption{Colour-magnitude diagram showing main-sequence stars and fixed-tempearture Dyson sphere models. The orange and red stars represent Sun-type and Proxima Centauri-type stars, respectively, while the grey dots represent other main-sequence stars. The coloured dots demonstrate how the Sun and Proxima Centauri shift their location in the diagram when we apply a DS model with $T_{\rm DS} = 300$ K and different covering factors.}
    \label{fig:ph_covering}
\end{figure}

\begin{figure}
	\includegraphics[width=\columnwidth]{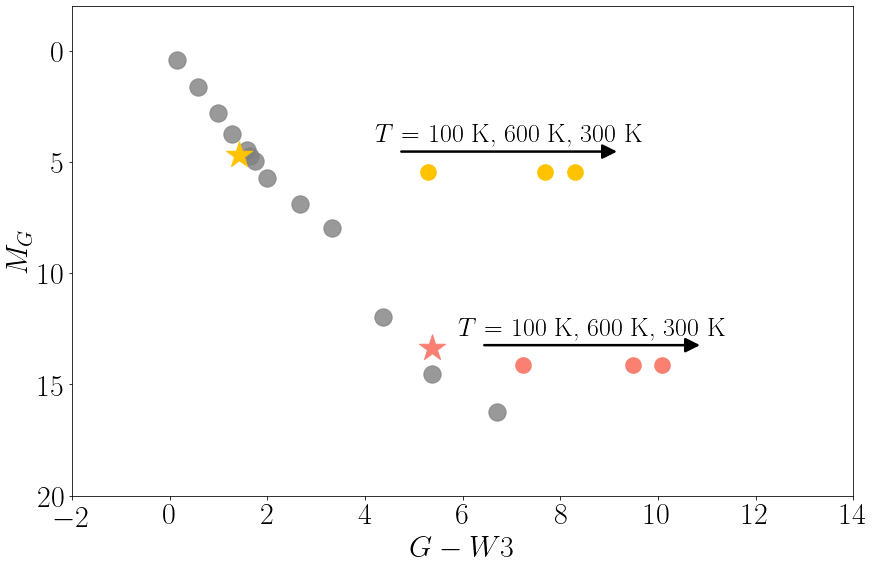}
    \caption{Colour-magnitude diagram showing main-sequence stars and Dyson sphere models with a fixed covering factor. The orange and red stars represent a Sun-type star and a Proxima Centauri-type star, while the grey dots are other main-sequence stars. The coloured dots demonstrate how the Sun and Proxima Centauri would be shifted in the diagram when applying a DS model with $\gamma = 0.5$ and different temperatures.}
    \label{fig:ph_temperature}
\end{figure}

In Figures~\ref{fig:ph_covering} and~\ref{fig:ph_temperature} we illustrate how the position of stars would change in the ($G-W3$,$G$) CMD due to presence of Dyson stars with various $\gamma$ and $T_{\rm DS}$. We use a Sun-like ($T = 5778$ K, L$_{\mathrm{bol}} = 1$ L$_{\odot}$) and a Proxima Centauri-like star ($T = 3042$ K, L$_{\mathrm{bol}} = 0.0017$ L$_{\odot}$) as examples. 


\subsubsection{Covering Factor}
\label{sec:covering_factors}
In Figure~\ref{fig:ph_covering}, we study how the covering factor affects our models for a fixed DS temperature of 300 K.  First, we notice how the brightness in the $G$ band decreases as $\gamma$ is increased. According to Equation~\ref{eq:mag}, the magnitude of the system in the $G$ band should be $G = -2.5\log(10^{-G_{\star}/2.5} + 10^{-G_{\rm DS}/2.5})$. However, a 300 K blackbody does not radiate in the optical at a comparable scale relative to stars. Because of this, the magnitude of the system becomes $G \approx G_{\star} = G_{\star,O} - 2.5\log(1-\gamma)$. As we can see, the drop in the optical depends only on the covering factor.



Another observation is that stars shift to redder regions (higher $G-W3$) of the colour-magnitude diagram as the covering factor increases. There are two reasons for this phenomenon. First of all, as we increase $\gamma$, we increase the drop in the optical flux, which translates as an increase in the $G$ magnitude, hence, a shift to redder regions. Second, we have an excess in the mid-infrared, which implies a decrease in the value of the $W3$ magnitude and consequently a further redward displacement. We know that the brightening of the flux in $W3$ must depend on the covering factor since a higher covering factor denotes a brighter DS by definition. Both phenomena cooperate in shifting the stars to the red regions of the colour-magnitude diagram.


\subsubsection{Temperature}
In Figure~\ref{fig:ph_temperature} we illustrate the effect in the ($G-W3$,$G$) CMD of changing the temperature for a DS with fixed covering factor. In this Figure, we model DS with a covering factor of 0.5 and temperature with values of 100, 300, and 600 K.


Since the covering factor is fixed, the drop in the optical is the same for every model, but every $G-W3$ colour shift is different because the DS temperature modulates the Planck function. In Figure~\ref{fig:ph_temperature}, we can see that the model that has the most significant shift is the one with a temperature of 300 K, for which the blackbody peaks in the $W3$ band. For the other two temperatures plotted, the peak occurs outside the $W3$ band which results in smaller $G-W3$ offsets from the main sequence.


\subsubsection{Additional remarks}
\label{sec:additional_remarks}

The analysis made in the previous sections was done by visualizing the behavior of the stars in a $G$, $G-W3$ colour-magnitude diagram. However, conclusions are general and apply to all combinations of Gaia-WISE colours for the range of temperatures adopted in this work.


Since we aim to derive upper limits based on WISE data, the range of DS temperatures considered in this paper is limited to 100 - 1000 K. DS operating outside this range \citep{Wright20} would not produce significant excess light in the mid-IR, but instead in the far-IR (for temperatures below 100 K) or in the near-IR (temperatures above 1000 K).


In Section~\ref{sec:covering_factors}, we neglected the flux contribution of the DS in the optical since we were analyzing a 300 K blackbody. We retain this assumption for the rest of the work since blackbodies in the range of temperatures adopted (100 K - 1000 K) do not contribute significantly to the Gaia $G_\mathrm{BP}$, $G$ or $G_\mathrm{RP}$ bands. Hence, we use Equation~\ref{eq:mstar} when we estimate how the $G$ magnitude of a star is affected by the DS.




Additionally, an important conclusion from the examples shown in Figures~\ref{fig:ph_covering} and~\ref{fig:ph_temperature} is how the colour shift produced by the DS depends on the colour of its host star. A DS constructed around a blue star results in a larger $G-W3$ shift than a DS with the same temperature and covering fraction built around a red star. This happens because a red star already has a low optical to mid-infrared flux ratio compared to a blue star, thereby resulting in a smaller relative shift in colour.


\subsection{Counting method}
\label{sec:counting_method}

To estimate the prevalence of DS in the Milky Way, we first identify the stars whose magnitudes and colours are consistent with those of DS models. For this purpose, we make use of CMDs that include the absolute magnitude in the $G$ band and the colours $G-W1$, $G-W2$, $G-W3$, and $G-W4$. We study the DS models in these diagrams to check if the stars in our samples fall in these regions for different combinations of temperatures and covering factors. This method is executed in two phases. The first parameterizes the boundaries between what is compatible with a DS and what is not. The second consists of checking if our stars fall in these regions.


\begin{figure*}
    \centering
    \includegraphics[width=0.95\textwidth]{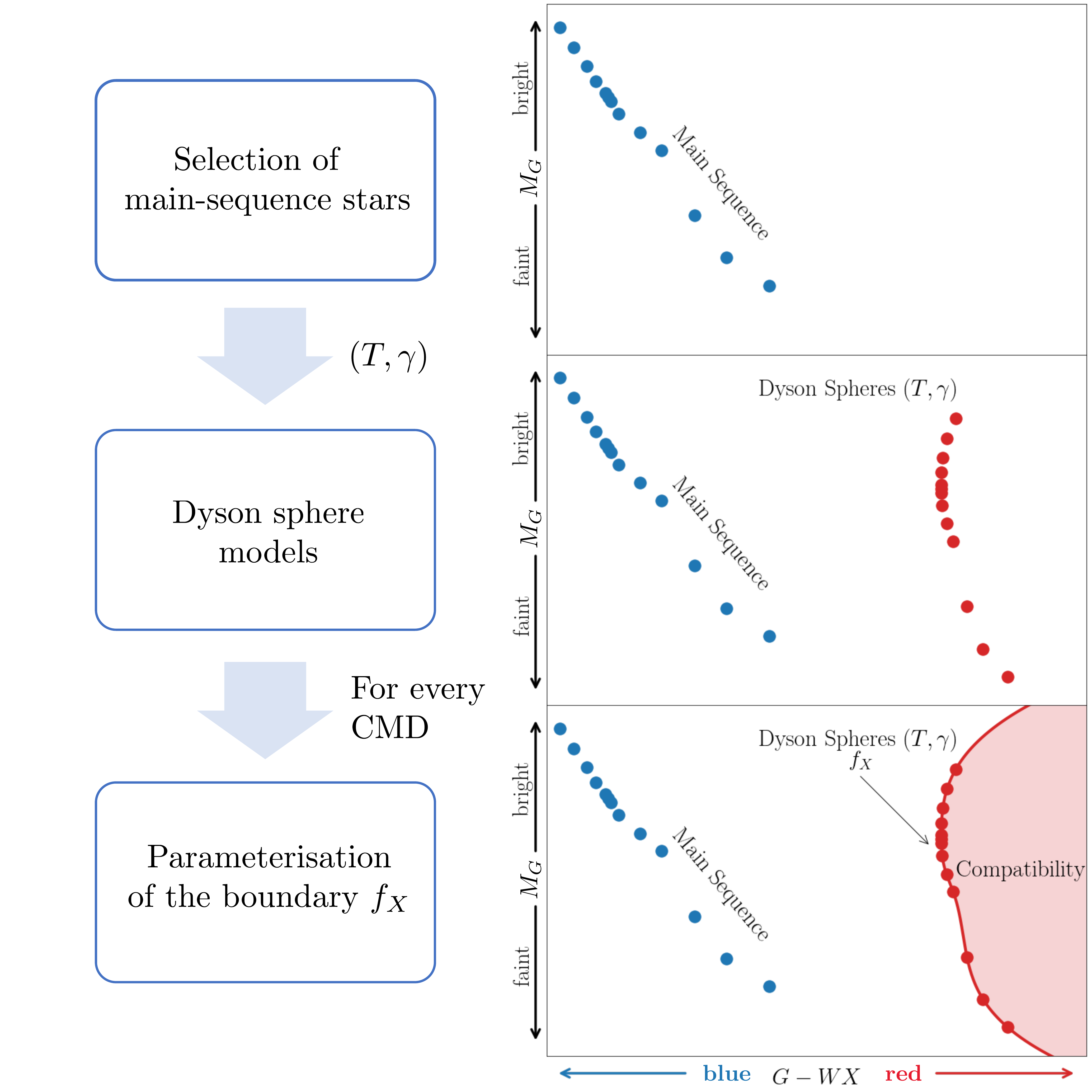}
    \caption{Summary of the first phase to count stars and DS candidates: parameterisation of boundaries. First, we select main-sequence stars, represented with blue dots. Then, we apply DS models of a given $T$ and $\gamma$ over the main-sequence stars, represented with red dots. We parameterise the boundary of the DS models with a mathematical function $f_X$, where $X$ is the number 1, 2, 3, or 4 to cover all the bands in the WISE mission.} 
    \label{fig:phase1}
\end{figure*}

The first phase is summarized in Figure~\ref{fig:phase1}. We select main-sequence stars in our sample of stars and then apply DS models of a given temperature and covering factor. Finally, we parameterise the boundary created by these models. For every WISE band ($WX$) used in the CMD, we get a different parametrisation of the border, where $f_X$ symbolizes the function that mathematically describes the boundary for WISE band $x$. For the purpose of this paper, we say that all the stars falling on the right side of this boundary are sources compatible with a DS of a given temperature and covering factor.

\begin{figure*}
    \centering
    \includegraphics[width=0.95\textwidth]{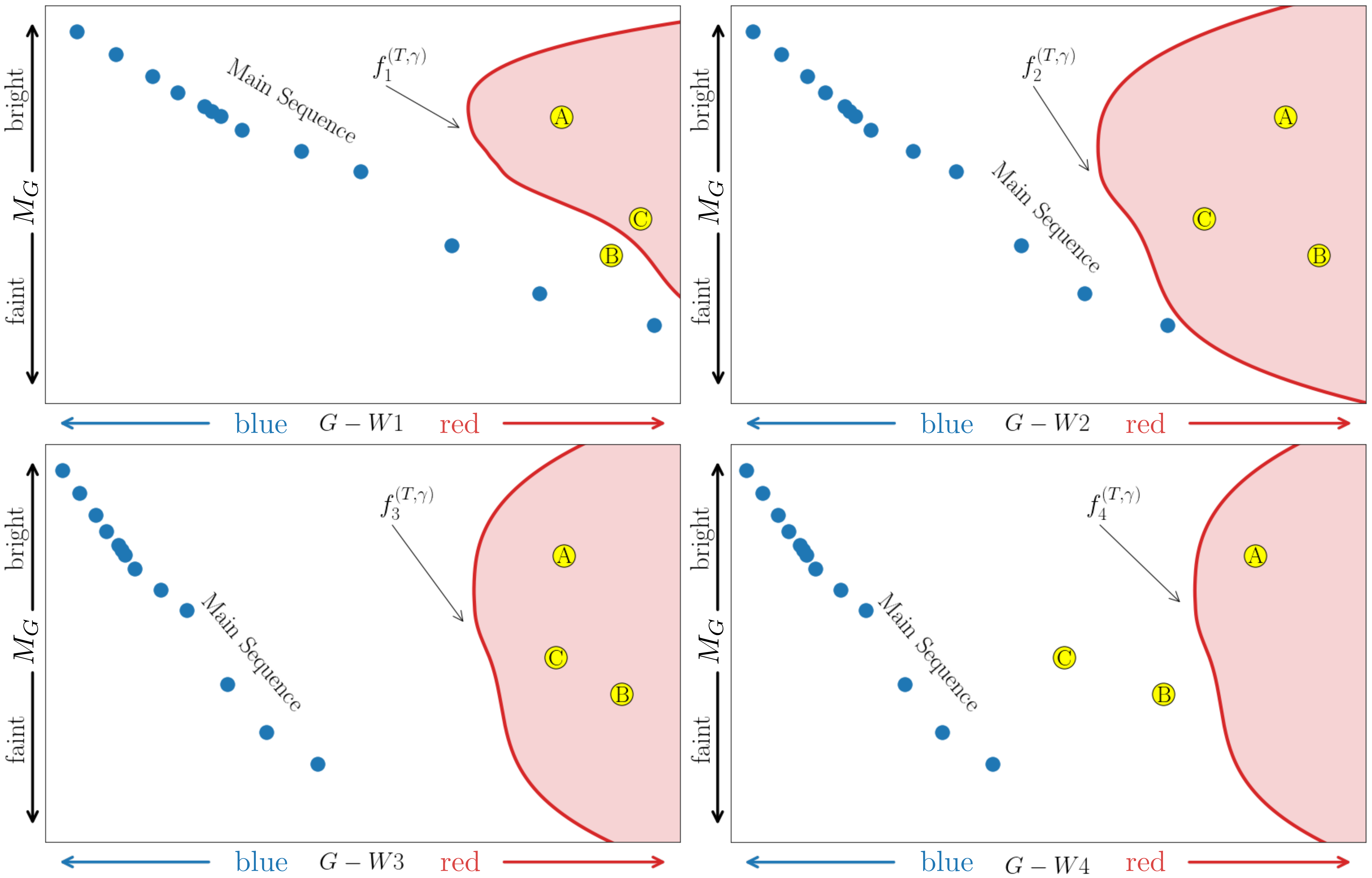}
    \caption{Summary of the second phase to count stars and DS candidates: location of stars in the compatibility regions for Dyson spheres. Here we show all the colour-magnitude diagrams in use. In these diagrams blue dots stand for main-sequence stars, while the red lines represent the boundaries of the compatibility regions which are represented with a parametrisation $f_1$, $f_2$, $f_3$, and $f_4$ for the diagrams $G-W1$, $G-W2$, $G-W3$, and $G-W4$, respectively.}
    \label{fig:procedure2}
\end{figure*}

The second phase consists of counting the stars that fall in the compatibility region of DS for all the combinations of colour-magnitude diagrams that we are using. This stage is depicted in Figure~\ref{fig:procedure2}. Here we see representative main-sequence stars and examples of stars exhibiting mid-infrared excesses. The boundary is parameterised by the functions $f_1$, $f_2$, $f_3$, and $f_4$ for the colours $G-W1$, $G-W2$, $G-W3$, $G-W4$, respectively. A star is only considered fully compatible with a DS of a given $T_\mathrm{DS}$ and $\gamma$ if it falls in the compatibility region for all the diagrams used. In Figure~\ref{fig:procedure2}, stars A, B, and C are examples that show how our criterion works. Star A falls in the compatibility region of all the diagrams used; hence, it represents a potential DS candidate for the $T_\mathrm{DS}$ and $\gamma$ used. On the other hand, star B does not fall in the compatibility region of the diagrams $G-W1$ and $G-W4$, so we dismiss it as a potentital DS. Similarly, star C does not fall in the compatibility region of the diagram $G-W4$. We discard these stars because the mechanism that produces excesses in specific bands cannot be the smooth blackbody continuum associated with the DS models for this combination of $T_\mathrm{DS}$ and $\gamma$.


Finally, the upper limit on the fraction of DS in the sample is obtained by dividing the number of Dyson sphere-like sources by the number of sources in the sample as described by: 

\begin{equation}
f_\mathrm{DS}(T_\mathrm{DS},\gamma)= \frac{N_\mathrm{DS}}{N_\mathrm{stars}(\gamma) + N_\mathrm{DS}},
\label{eq:fraction}
\end{equation} where $f_\mathrm{DS}$ is the fraction of stars that resemble DS for a given $T_\mathrm{DS}$ and $\gamma$, $N_\mathrm{DS}$ is the number of stars that fall in the DS region of the CMDs, and $N_\mathrm{stars}$ is the number of stars that do not resemble DS.


An important point to consider when estimating our fractions turns out to be the number of stars ($N_\mathrm{stars}(\gamma)$) in the denominator of Equation~\ref{eq:fraction}. The instrumental detection limit naturally reduces the number of stars that can produce a detectable model of a certain $\gamma$. As an example, if the $G$-band apparent magnitude detection limit is $m_\mathrm{limit}$, then detectable DS models can only be produced by stars with apparent magnitude $m_G<m_\mathrm{limit} + 2.5\log(1-\gamma)$, whereas fainter stars will have DS models too faint to be detected. We take this phenomenon into account when estimating the fraction of stars that are compatible with DS.

Additionally, we notice that some stars lack detections in the WISE filters. For a star below the WISE detection threshold, we use the $S/N<2$ WISE upper limits on its flux, rather than the flux itself, when assessing whether it falls inside or outside a given DS compatability region in the CMDs.

\subsection{First sample}
\label{sec:data_samples}

We use combined Gaia DR2 - AllWISE dataset for this work. The cross-match between Gaia DR2 and AllWISE sources is done by using the \textit{AllWISE best neighbour} catalog provided by Gaia. We use as many stars as possible, therefore our samples are just limited by the instruments, especially Gaia, that has a limiting magnitude $G = 21$. Since distances based on simply inverting the parallax can lead to highly spurious results in cases where the parallax uncertainty is high, we instead adopt the distances provided by \citet{BJ}, which are based on the Gaia DR2 parallaxes and a prior that depends on Galactic longitude and latitude. 

As a first approach to estimating upper limits on the fraction of stars that could potentially host DS, we select a sample of stars within 100 pc in Gaia DR2 + AllWISE. In total 265724 sources are located within 100 pc. Then, we use the method described in Section~\ref{sec:counting_method} to estimate the number of sources compatible with DS for different $T_\mathrm{DS}$ and $\gamma$. Since these stars are in the local bubble, we assume zero extinction (the effect of extinction is further discussed in Section~\ref{sec:extinction})




Then, we proceed to select the main-sequence stars on which to base the DS models for this sample. Main-sequence stars are selected using their absolute $G$ magnitude and $G_\mathrm{BP} - G_\mathrm{RP}$ colour. Objects lacking this colour measurements are removed, but this affects only $\approx 0.8$\% of the objects in the 100 pc sample.

\begin{figure}
    \centering
    \includegraphics[width=\columnwidth]{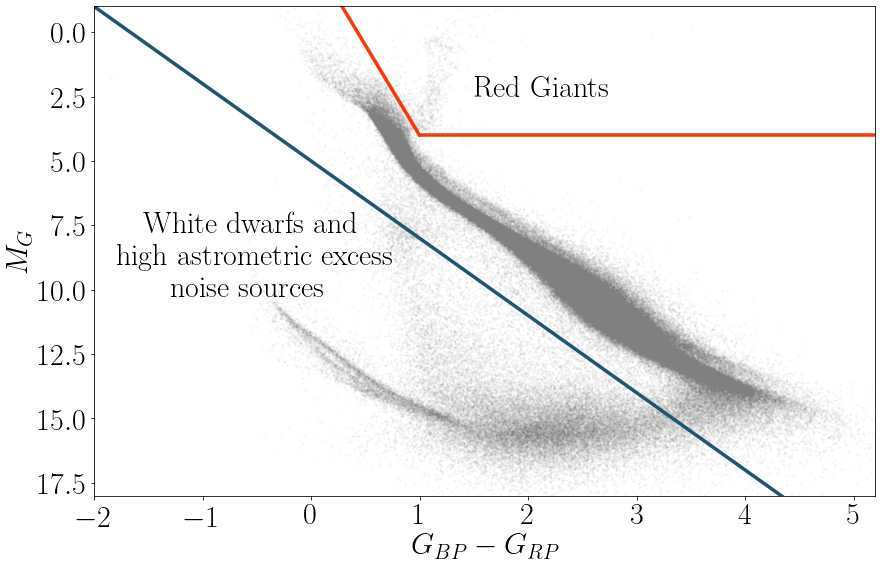}
    \caption{Stars in the Gaia DR2 + AllWISE sample within 100 pc and the cuts we use. In this figure all the stars in the sample are depicted with grey dots. We use the red line delimits main-sequence stars from red giants, while we use the blue line separates white dwarfs and high astrometric excess noise stars from main-sequence stars.}
    \label{fig:ms}
\end{figure}

We illustrate the selection of main-sequence stars in Figure~\ref{fig:ms}, which features a colour-magnitude diagram with colour $G_\mathrm{BP} - G_\mathrm{RP}$ vs. absolute $G$ magnitude, and lines that demonstrate the pruning of white dwarfs, objects with high astrometric excess noise and post main-sequence stars. High astrometric excess noise sources are stars for which the astrometric solution has failed and/or stars in multiple systems for which the single-star solution does not work \citep{gaia_hr}. These stars correspond to the cloud between the main-sequence and the white dwarf sequence. 

The functional form used to filter red giants corresponds to Equation~\ref{eq:postms}, whereas the function used to filter white dwarfs and high astrometric excess noise is given by Equation~\ref{eq:wd}. In these equations, $M_{G}$ corresponds to the absolute magnitude of the star in the $G$ band and $G_\mathrm{BP} - G_\mathrm{RP}$ its colour. 

\begin{equation}
    M_{G}\text{ }<\text{ }4 \text{ and } M_G < 7\cdot(G_\mathrm{BP} - G_\mathrm{RP}) - 3
    \label{eq:postms}
\end{equation}

\begin{equation}
    M_G > 3\cdot(G_\mathrm{BP} - G_\mathrm{RP}) + 5
    \label{eq:wd}
\end{equation}

Even though we attempt to reject post main-sequence objects and sources with high astrometric excess noise, these cuts are admittedly not perfect. Nevertheless, we show in Appendix~\ref{sec:pms_effect} that residual members from these classes of objects do not contaminate our models significantly.



\subsubsection{Models and their behavior}
\label{sec:behavior}

Once all main-sequence stars are selected, we proceed to apply DS models to these. However, the bolometric luminosity of each star is required as an input parameter, and Gaia DR2 only provides estimations for a few of them \citep{gaia_apsis}. We estimate the missing luminosities by using a polynomial fitting function where we extrapolate from their absolute magnitudes in the $G$ band. This can be done since we are working with main-sequence stars. The details on the luminosity fitting function used are in Appendix~\ref{sec:lum}.

\begin{figure*}
    \centering
    \includegraphics[width=0.95\textwidth]{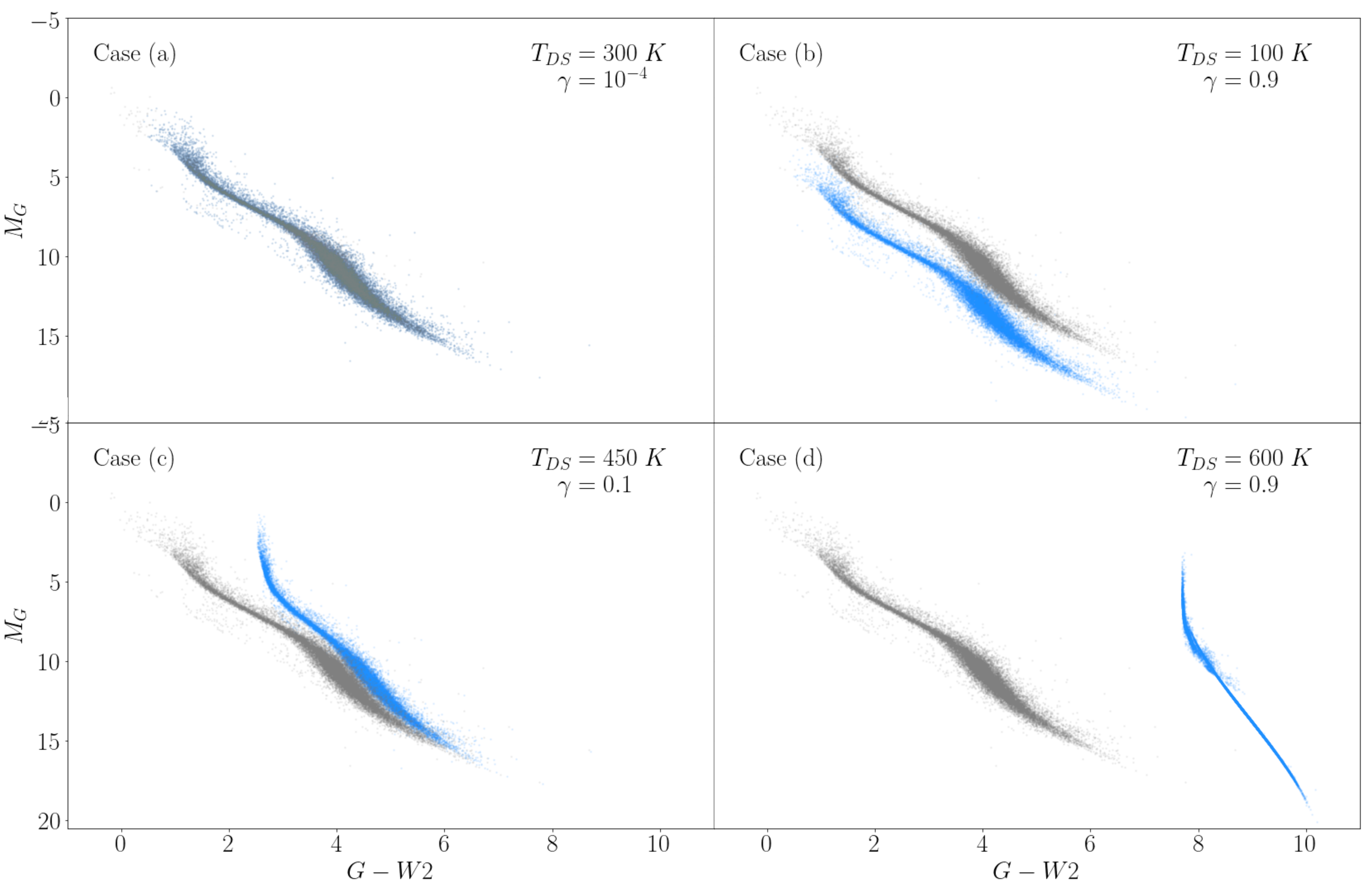}
    \caption{Examples of how our models behave in colour-magnitude diagrams for different combinations of DS temperature and covering factor. Main-sequence stars are depicted with grey dots, while blue dots correspond to our DS models. The upper-left panel represents the low covering factor case, where DS are indistinguishable from the MS. The upper-right panel shows the behavior of cool DS. Only the optical drop is recovered for colours using the $W1$ and $W2$ bands. The lower-left panel depicts the intermediate case, when some DS separate from the MS, and some do not. The lower-right panel show the ideal case, when all DS are clearly distinguishable from the main sequence.}
    \label{fig:cmd_models}
\end{figure*}

With all the bolometric luminosities, we apply our DS models to all the stars in the main-sequence sample. We do this for a range of temperatures between 100 and 1000 K and covering factors from $10^{-4}$ to $0.9$. In Figure~\ref{fig:cmd_models}, we show some examples of our models in CMDs of $G-W2$ vs. $M_G$. We identify four trends, which we describe in more detail below.


\paragraph*{Case a - Low covering factor}

First of all, we have the case of DS with low covering factors, which is illustrated in the upper-left panel of Figure~\ref{fig:cmd_models}. In this example, the models have a temperature of 300 K and a covering factor of $10^{-4}$. We can see how both main-sequence stars and models overlap almost completely. For a covering factor this small, both the optical dimming and the mid-IR excess are too small to be detected. Consequently, DS of this type are not distinguishable from normal main-sequence stars in the present data set.


\paragraph*{Case b - Cool Dyson spheres}
\label{sec:cool_dyson}

In the upper-right panel of Figure~\ref{fig:cmd_models}, we show models with a temperature of 100 K and a covering factor of $0.9$. The models behave like the normal main sequence but shift downwards for this particular choice of colour ($G-W2$). This phenomenon happens because the blackbody spectrum for a 100 K DS gives a negligible contribution to the W2 band, so the only signature recovered is the drop in the optical flux. This phenomenon happens only for cool DS when we try to use the colours $G-W1$ and $G-W2$ to measure the infrared excess.


The same effect is not seen for hot DS ($T_{\rm DS} > 600$ K)  when $G-W3$ and $G-W4$ are used as a measure of infrared excess. Even though the blackbody spectrum for hot DS peaks at wavelenghts longward of the $W3$ and $W4$ passbands, the tail of their spectra still contributes significantly at these wavelengths (as seen in Figure~\ref{fig:spectra}).


\paragraph*{Case c - Intermediate cases}
In the lower-left panel of Figure~\ref{fig:cmd_models}, we have an example of an intermediate case, when the DS models are only significantly offset from the main sequence in the case of the brightest stars. In this example, the DS temperature is 450 K, and the covering factor is 0.1. The phenomenon described in Section~\ref{sec:additional_remarks} explains why these DS models are more clearly separated from the main sequence for brighter stars. In general, the range of magnitudes for which the DS are disinguishable from the main sequence depends both on the DS parameters ($T,\gamma$) and on the colour used.


\paragraph*{Case d - Ideal cases}
Finally, in the lower-right panel of Figure~\ref{fig:cmd_models}, we illustrate how the models behave for a DS with a temperature of 600 K and a covering factor of 0.9. In this case, all DS models are clearly separated from the main sequence. This happens when the DS blackbody peak falls in the wavelength range of the WISE bands, and the covering factor is so high that all DSs give rise to a significant colour shift, regardless of whether they are hosted by bright (hot, blue) or faint (cool, red) main-sequence stars. 


\subsubsection{Boundaries and counting}
\label{sec:counting_stars}
\begin{figure}
    \centering
    \includegraphics[width=\columnwidth]{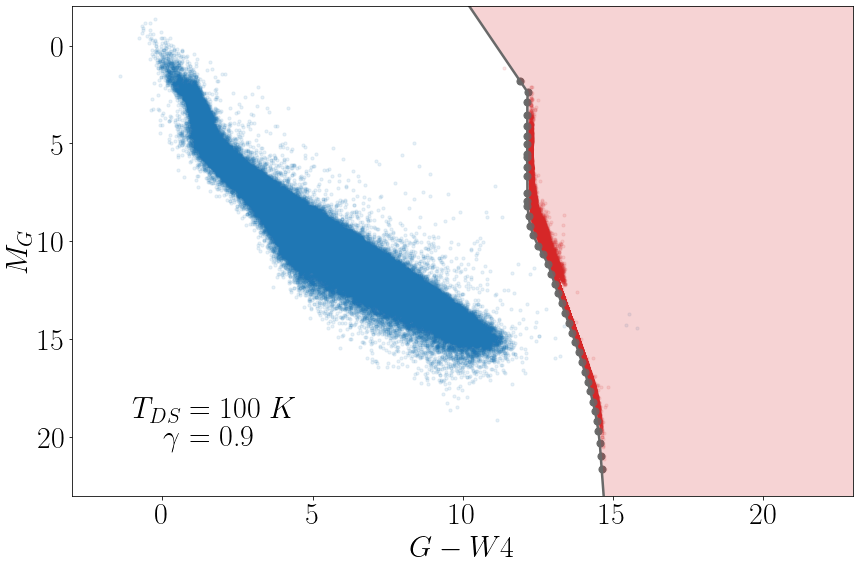}
    \caption{Colour-magnitude diagram that illustrates the border between main-sequence stars and sources compatible with Dyson spheres. In this diagram we show the distribution of stars (blue dots) and the models (red dots) for a DS with $T_{\rm DS} = 100$ K and $\gamma = 0.9$. The black dots represent the leftmost stars chosen every 0.5 G magnitude bin that are used to create the boundaries by using a linear interpolator. The red region represents the compatibility region for this model in this specific colour-magnitude diagram.}
    \label{fig:region}
\end{figure}

To estimate the number of objects potentially compatible with DS within a sample, we define four boundaries (one for each of the colours $G-W1$, $G-W2$, $G-W3$ and $G-W4$) between the DS models and the main-sequence stars. Figure~\ref{fig:region} shows an example of how these boundaries are established. For a given $T_{\rm DS}$ and $\gamma$ we derive the DS models (red dots) and pick the leftmost objects among these models for every $\Delta(G)=0.5$ mag bin (black circles). With these objects, we determine the border by using a linear interpolator (black line) and extrapolate this line for values beyond the models we have. This procedure is carried out for every colour-magnitude diagram independently.


With every boundary defined, we proceed to count the number of stars on either side. However, objects are not considered DS candidates unless they fall in the DS regions of all four colour-magnitude diagrams considered (including $G-W1,G-W2,G-W3$, and $G-W4$ on the x-axis). 

\section{Results}
\label{sec:results}
\subsection{Exclusion maps for the 100 pc sample}
\begin{figure}
    \centering
    \begin{subfigure}[b]{0.48\textwidth}
    \centering
    \includegraphics[width=\textwidth]{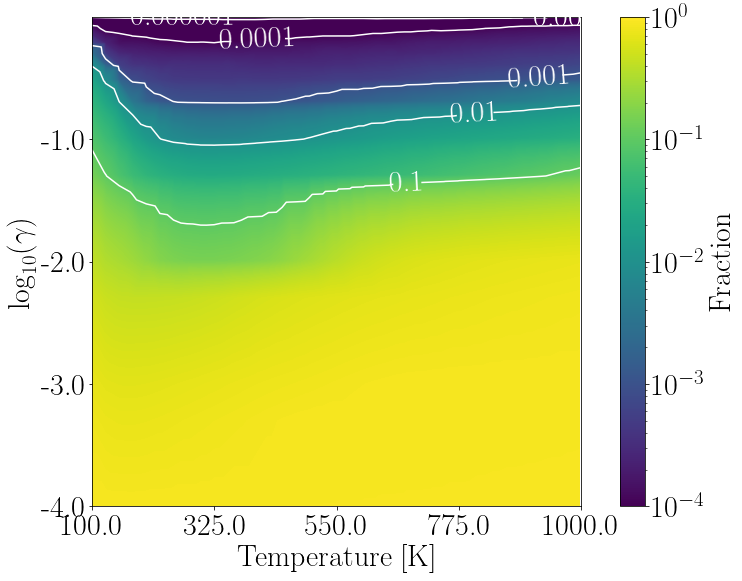}
    \end{subfigure}
    \begin{subfigure}[b]{0.48\textwidth}
    \centering
    \includegraphics[width=\textwidth]{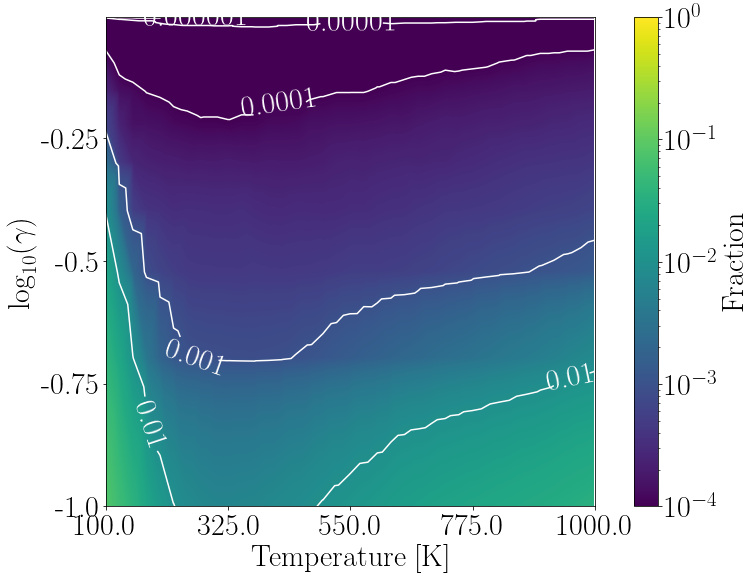}
    \end{subfigure}
    \caption{Exclusion maps for stars within 100 pc: Colour maps that represent the fraction of stars that are compatible with DS of a given temperature and a covering factor. In the top panel we show the exclusion maps for temperatures between 100 and 1000 K, and covering factors between $10^{-4}$ and 0.9. In the bottom panel we show the same exclusion map, but for covering factors between 0.1 and 0.9.}
    \label{fig:exc}
\end{figure}

Using the computational machinery outlined in Section~\ref{sec:counting_method} and Section~\ref{sec:counting_stars}, we now set out to derive upper limits on the fraction of objects that could potentially host DS within 100 pc in the Gaia DR2 + AllWISE combined dataset. Since the CMD boundaries for DS candidates depend on the assumed DS temperatures and covering factors, so will the limits. By considering temperatures from 100 to 1000 K and covering factors from 10$^{-4}$ to 0.9, we arrive at the exclusion map shown in Figure~\ref{fig:exc}.

 From this exclusion map, we can recognize the trends already discussed in Section~\ref{sec:behavior}. For covering factors below $\log_{10}(\gamma) = -2$, the upper limit on the fraction of potential DS is close to unity, i.e., a very weak limit. This means that we are unable to set meaningful limits on such low-$\gamma$ DS. In the case of covering factors between $\log_{10}(\gamma) = -2$ and $\log_{10}(\gamma) = -0.25$, the upper limits become progressively stronger, as the DS models for the brighter stars in the sample detach from the main sequence and the mid-IR excess becomes recognizable. For covering factors above $\log_{10}(\gamma) \approx -0.125$ ($\gamma$ $\approx$ 0.75), the upper limits are below 10$^{-4}$ for all temperatures considered.

\subsection{Exclusion maps for stars out to 5 kpc}

\begin{figure*}
    \centering
    \includegraphics[width=\textwidth]{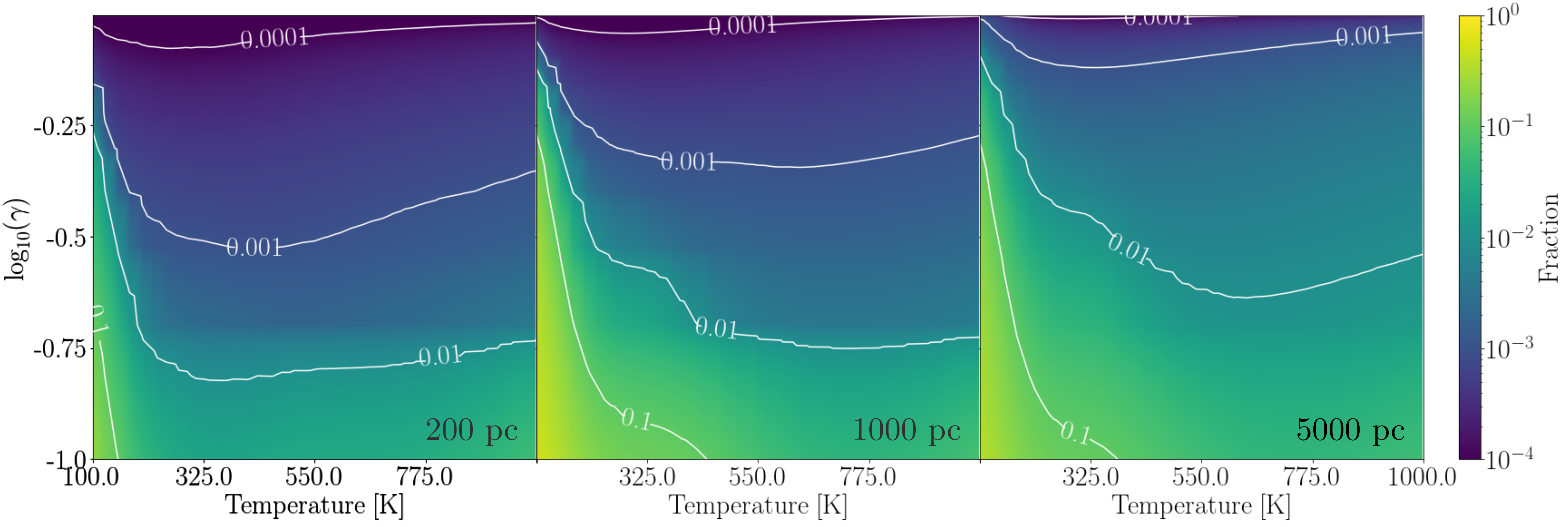}
    \caption{Exclusion maps at various maximum distances: Colour maps that represent the fraction of stars that are compatible with DS of a given temperature and a covering factor. In the left, middle, and right panels we show the exclusion maps for the sample of stars within 200, 1000, and 5000 pc respectevely.}
    \label{fig:everything}
\end{figure*}

In Figure~\ref{fig:everything} we show exclusion maps for the samples containing stars at distances up to 200, 1000, and 5000 pc. As the effects of interstellar dust reddening become more severe at large distances, objects exhibit more scatter in the redward direction across the CMD, which makes the main sequence more difficult to isolate. Instead of using Equations~\ref{eq:postms} and ~\ref{eq:wd} to filter out stars not located on the main sequence, we have in the 5000 pc sample resorted to Equation~\ref{eq:random} to select main-sequence candidates on which our DS models are based. 
\begin{equation}
    M_G \leq 1.5\cdot (G_\mathrm{BP} - G_\mathrm{RP}) - 0.1
    \label{eq:random}
\end{equation}

In Figure~\ref{fig:everything}, we limit the constraints plotted to covering factors above 0.1, since the upper limits derived for lower $\gamma$ are very weak. Overall, the upper limits become progressively weaker when the maximum distance limit of the sample is increased. 

The primary reason for this is that the fraction of Gaia stars that fall above the detection threshold of WISE drops with distance. At large distances, intrinsically faint, red main-sequence stars tend to drop out of the Gaia sample, which leaves a larger fraction of intrinsically brighter, bluer main sequence stars.  Because of this, the fraction of stars that lie just above the Gaia detection threshold, yet below the WISE detection threshold, grows with increasing distance.

As all the large star-forming regions of the Milky Way are located at distances beyond 100 pc, a larger number of Dyson-sphere interlopers in the form of Young Stellar Objects (YSOs) also start entering the sample. Finally, samples featuring objects at larger distances (smaller parallaxes) will also include more objects with highly uncertain distance estimates, and hence highly unreliable absolute magnitudes. This includes intrinsically very bright objects from outside the distance cut-off (even extragalactic objects, like quasars) contaminating the samples. We stress that none of these samples are complete in a volumetric sense, and instead flux-limited by the Gaia DR2 and AllWISE detection thresholds.

Table~\ref{tab:table} summarizes the DS constraints set by each sample for the case of $T_{\rm DS} = 300$ K and $\gamma = 0.1, 0.5, 0.9$. As seen, the 100 pc sample gives the strongest constraints, and the 5000 pc the weakest. When comparing the upper limits on the Dyson sphere fraction for the case of $\gamma = 0.9$ DS, we note that the upper limits are weakened by about one order of magnitude as the sample size is increases by three orders of magnitude ($\sim 2.7\times 10^5$ objects compared to $\sim 2.9\times 10^8$ objects, when going from a maximum distance of 100 pc to and 5000 pc).




\begin{table*}
    \centering
    \begin{tabular}{c|c|c|c|c}
    \hline
    Radius [pc] & Size sample & \multicolumn{3}{ c| }{Fraction of stars compatible with a DS with $T_{\rm DS} = 300$ K} \\
    & & $\gamma = 0.1$ &  $\gamma$ = 0.5 & $\gamma$ = 0.9 \\
    \hline
        100 & 265724 & 6.6$\times$10$^{-3}$ & 1.9$\times$10$^{-4}$ & 1.8$\times$10$^{-5}$\\
        200 & 1847472 & 2.2$\times$10$^{-2}$ & 4.7$\times$10$^{-4}$ & 7.4$\times$10$^{-5}$ \\
        300 & 5243957 & 4.8$\times$10$^{-2}$ & 4.8$\times$10$^{-4}$ & 8.6$\times$10$^{-5}$\\
        400 & 10384485 & 6.7$\times$10$^{-2}$ & 5.6$\times$10$^{-4}$ & 1.1$\times$10$^{-4}$\\
        500 & 16831302 & 8.9$\times$10$^{-2}$ & 7.5$\times$10$^{-4}$ & 1.9$\times$10$^{-4}$\\
        600 & 24162082 & 1.0$\times$10$^{-1}$ & 7.5$\times$10$^{-4}$ & 1.9$\times$10$^{-4}$\\
        700 & 32123156 & 1.1$\times$10$^{-1}$ & 7.8$\times$10$^{-4}$ & 1.9$\times$10$^{-4}$\\
        800 & 40575873 & 1.2$\times$10$^{-1}$ & 8.5$\times$10$^{-4}$ & 2.0$\times$10$^{-4}$\\
        900 & 49472732 & 1.2$\times$10$^{-1}$ & 9.4$\times$10$^{-4}$ & 2.0$\times$10$^{-4}$\\
        1000 & 58769015 & 1.3$\times$10$^{-1}$ & 1.0$\times$10$^{-3}$ & 2.1$\times$10$^{-4}$ \\
        5000 & $\sim$2.9$\times$10$^{8}$ & 1.6$\times$10$^{-1}$ & 7.1$\times$10$^{-3}$ & 8.3$\times$10$^{-4}$\\
        \hline
    \end{tabular}
        \caption{Summary of the samples analyzed. In this table we summarize the samples analyzed in this work. We show the distance considered in each sample, the size of the sample, and the percentage of sources compatible with a DS of $T = 300 $ K and covering factors 0.1 , 0.5, and 0.9.
        }
    \label{tab:table}
\end{table*}

We stress that, when counting the sources that fall in given DS region for a certain $\gamma_0$, we include all the stars on the right-hand side of the boundary in the CMD. This means that we are selecting stars consistent with $\gamma$ > $\gamma_0$ (for a fixed $T_{\rm DS}$). However, this $\gamma$ cannot reach unity, since the DS would then completely obscures these sources and render it undetectable in the optical. 

\section{Discussion}
\label{sec:discussion}

\subsection{Previous upper limits}

Prior to the current study, the most recent work on upper limits was that of \citet{Carrigan09}. As a preliminary part of his search, Carrigan studied the photometry provided by the IRAS mission. Based on the fluxes at 12, 25, and 60 $\mu$m, he concluded that less than one in 10000 of the $\approx 2.5\times 10^5$ IRAS sources could be DS, in the sense of being consistent with pure black-body spectra in the 150 K to 500 K range across these IRAS bands. This is, for instance, similar in terms of sample size and limit for our 100 pc sample in the case of $\gamma>0.6$ at $T_\mathrm{DS}=300$ K. Carrigan also searched for DS by analyzing the atlas compiled by the Calgary group. This sample contained information from the IRAS low-resolution spectrometer for $\approx 1.1\times 10^4$ sources. After applying different cuts, he ended up with 16 objects called 'somewhat interesting sources.' However, he focused on pure DS, i.e., $\gamma = 1$. This assumption implies pure thermal blackbody emission from the DS and complete obscuration of the host starlight. Unlike Carrigan's work, here we focus on the behavior of partial DS, where the optical flux of the star remains relevant. Moreover, our method for estimating upper limits relies on the location of stars in different CMDs, whereas Carrigan focused on temperature estimations. The most significant difference between our upper limit estimates turns out to be the characteristics of the dataset used, as our largest dataset is roughly three orders of magnitude larger and also includes distance estimates for the sources. 



\subsection{Feedback on the host star}
\citet{Huston21} recently highlighted the importance of the feedback from DS on the properties of their host stars. They found that the interiors of low-mass stars may be affected under the presence of high feedback values, whereas more massive stars ($\geq$ 1 M$_{\odot}$) remain insensitive to internal changes. While our models neglect the impact of feedback on the host stars, the DS temperatures considered in our work correspond to DS radii much larger than the range for which such feedback would have detectable consequences. As an example, let's take a solar-type star and the highest temperature used in our work (1000 K) with a covering factor of 0.90. The feedback factor, or percentage of the star's luminosity reflected back on itself, is 0.04\%. For a 1\% change in a solar-type star's nuclear luminosity, the feedback factor must be at least 45\%, and for a 1\% change in a solar-type star's effective temperature, a feedback factor of roughly 5\%. Even for lower-mass stars which are more strongly affected by feedback, a feedback level of roughly 1\% is required in order to change nuclear luminosity by 1\%. The effective temperatures of these stars are not significantly affected for any feedback levels below 50\%. Thus, the non-reflective and relatively cool DS models in this work would not produce feedback strong enough to cause noticeable changes in stellar properties.

\subsection{Multi-layer Dyson spheres}
Many variations on the DS theme have been proposed since the seminal paper by \citet{dyson60}. One of them consists of many DS layers operating at different temperatures. In the case of such objects, our models would still apply for some specific configurations while potentially failing for others. For instance, in a two-layer system, the outer layer could absorb some of the thermal light emitted from the inner sphere. If the radiation of the highest-$\gamma$ layer completely dominates over the other one, our limits would roughly correspond to the properties ($T_\mathrm{DS}$,$\gamma$) of that highest-$\gamma$ layer. However, in cases where both layers have similar contributions, the situation becomes more complicated. As an example, a double-layer system where the two layers operate at  $T_{\rm DS} = 275$ and $300$ K respectively, and each have $\gamma = 0.7$, would result in the same number of compatible sources and upper limits as a single layer with $T_{\rm DS} = 300$ and $\gamma = 0.9$. A double-layer system with $T_{\rm DS} = 675$ and $700$ K, with each layer at $\gamma = 0.781$ also produces the same result.

\subsection{Dyson sphere interlopers}
The WISE mission is sensitive to the expected mid-infrared signatures of DS in the 100-1000 K range, but dusty sources can produce similar signatures. PAH emission features fall in the 3.4 and 12 $\mu m$ filters. Additionally, the 4.6 $\mu m$ filter measures the continuum emission from tiny grains. The 22 $\mu m$ filter sees both stochastic emission from small grains and the tail of thermal emission from large grains \citep{wright_wise}. Hence, sources surrounded by thick dust shells can exhibit significant mid-IR fluxes and populate the same regions of Gaia-WISE CMDs as our DS models. While bona fide DS are expected to produce a smooth mid-IR continuum, our method of counting sources in separate CMDs does not distinguish between sources with emission peaks in the separate WISE filters and those for which the WISE colours reflect a pure black-body spectrum. Young stellar objects (YSOs) surrounded by dusty cocoons and evolved stars ejecting dust into the interstellar medium are therefore significant DS interlopers in the current study and likely dominate our upper limits on the fraction of objects exhibiting Dyson sphere-like characteristics. A spectroscopic analysis would be necessary to uncover the true nature of individual DS candidates, but a more detailed analysis of the spectral energy distribution from the optical to the far-IR can also help weed out many types of dusty objects. 

To get a sense of the relevance of interlopers in our analysis, we have examined the auxlilary data available for the 5 objects at distances out to 100 pc that in our diagnostic CMDs are deemed compatible with DS characterized by $T_{\rm DS} = 300$ K and $\gamma = 0.9$. The available 2MASS photometry reveal that all of these have very red optical-NIR colours that indicate that they cannot be main sequence objects, unless they happen to be subject to several magnitudes of optical extinction due to interstellar dust. The latter scenario is implausible for stars within 100 pc \citep[e.g.,][]{Lallement19}, but it should be noted that these objects have highly uncertain distances \citep[as is evident from significantly reduced parallaxes for four out of the five objects in Gaia EDR3 compared to DR2;][]{Gaia21} and may in reality lie much further away. In either case, they are not good candidates for DS of the type we are considering in this paper. 

All of the five objects are located in nebular surroundings, as revealed by available mid-IR and$/$or far-IR images, indicating that they are likely to be YSOs associated with ongoing star formation. As there are no large star-forming regions known within 100 pc, this highlights the problem of contamination by more distant objects in the sample. For instance, one of these objects lies in the direct vicinity of (and is hence likely associated with) the Taurus molecular cloud, one of the nearest star-forming regions, at $\approx 140$ pc. For the four objects with EDR3 parallaxes, the distance estimates presented by \citet{Bailer-Jones21} also place them far beyond the 100 pc limit.

In the sample of stars within 100 and 200 pc, 110 sources are compatible with our $T_{\rm DS} = 300$ K, $\gamma = 0.9$ DS models. Here we resort to the classifications available in the SIMBAD database, and bin all the possible labels into 5
categories: pre-main-sequence stars, stars, post-main-sequence stars,
other types, and unclassified. The label 'stars' refers to sources that are simply classified as such, with no additional information about their type or
evolutionary stage according to SIMBAD. Thirty sources (27.3 \%) are classified as pre-main sequence stars, eight sources (7.2 \%) are classified as stars, five sources (4.5 \%) correspond to post main-sequence stars, and eight sources (7.2 \%) fall in the 'other' category. We note that most of these sources are uncategorized, and this trend becomes more pronounced for samples at even larger distances.

\subsection{Temporal variations}
Dyson spheres are usually envisioned as a swarm of small panels or satellites collecting energy from their host star. When hypothetically observing a star surrounded by a partial DS, the number of orbiting panels covering the stellar disk would fluctuate over time, which could give rise to temporal variations in the amount of unobscured star light emitted in our direction. However, whether such effects would be detectable depends on the sizes of the panels used. \citep{wright16} states that for a vast number of small objects, the array of satellites might appear as a translucent screen that cannot be easily detected. Intermediate-size panels would generate time variations that can be mistaken for other phenomena like photospheric noise or asteroseismic variations. Even larger objects/panels might generate light curves characterized by aperiodic events of almost arbitrary depth, duration, and complexity. Hence, while variability could provide an interesting auxiliary diagnostic, we cannot easily dismiss DS candidates based on whether they display variability. Light curve information should however be able to weed out some types of interlopers, like MIRA variables, that become interlopers when their distances are underestimated.



\section{Conclusions}
\label{sec:conclusions}
By analyzing the $2.9\times 10^8$ objects in the combined Gaia DR2 + AllWISE dataset samples, we have derived a set of conservative upper limits on the fraction of stars that could potentially host near-complete DS with covering fractions in the 0.1--0.9 range. The present analysis is limited to non-reflective DS operating at temperatures from 100--1000 K. 

Our conclusions can be summarized as follows:
\begin{enumerate}
\item The constraints tend to be the strongest for DS temperatures around 300 K (similar to the objects originally envisioned in \citealt{dyson60}), and become progressively weaker at both higher and lower temperatures (Figures~\ref{fig:ph_temperature}, ~\ref{fig:exc}, ~\ref{fig:everything}).
\item Among the $\approx 2.6\times 10^5$ stars located within 100 pc in the sample, less than one per $\sim 10^3$ objects is compatible with a 300 K DS with a covering factor of $\gtrsim 0.1$, less than one per $\sim 10^4$ sources with a covering factor $\gtrsim 0.5$, and one per $\sim 10^5$ with a covering factor of $\gtrsim 0.9$ (Figure~\ref{fig:exc}, Table~\ref{tab:table}).
\item Upper limits derived at distances beyond 100 pc are weaker, as objects with significant mid-IR excess may still lie below the detection threshold of WISE. Among the $\approx 2.9\times 10^8$ sources in the sample out to 5000 pc, as many as one per $\approx 3000$ objects could in principle be consistent with a 300 K, covering fraction 0.9 DS (Table~\ref{tab:table}, Figure~\ref{fig:everything}).
\end{enumerate}

In mid-2022, Gaia DR3 will provide stellar parameters deduced from spectral analysis, which will allow stronger constraints on the nature of the stars with mid-IR excesses associated with DS.

\section*{Acknowledgements}
MS acknowledges funding from the Royal Swedish Academy of Sciences. EZ acknowledges funding from the Magnus Bergvall Foundation and a sabbatical fellowship from AI4Research at Uppsala University. AK acknowledges funding from the Swedish National Space Agency (SNSA).

This work presents results from the European Space Agency (ESA) space mission Gaia. Gaia data are being processed by the Gaia Data Processing and Analysis Consortium (DPAC). Funding for the DPAC is provided by national institutions, in particular the institutions participating in the Gaia MultiLateral Agreement (MLA). This publication makes use of data products from the Wide-field Infrared Survey Explorer, which is a joint project of the University of California, Los Angeles, and the Jet Propulsion Laboratory/California Institute of Technology, and NEOWISE, which is a project of the Jet Propulsion Laboratory/California Institute of Technology. WISE and NEOWISE are funded by the National Aeronautics and Space Administration.

\section*{Data availability}

The data underlying this article will be shared on reasonable request to the corresponding author.





\bibliographystyle{mnras}
\bibliography{example} 




\appendix

\section{Post Main-Sequence and high astrometric excess noise sources}
\label{sec:pms_effect}

The process of creating DS models for main-sequence stars used in this paper involves an attempt to first filter out post main-sequence and high astrometric excess noise sources from the sample. However, since it is unlikely that this procedure removes every single object of this type, a small fraction of the DS models generated may in fact end up being based on stars that are not on the main sequence. Here, we explore what effects this may have on our overall result, using the 100 pc sample from Section~\ref{sec:data_samples} as a reference. 

First, we study the effect of white dwarfs (WD) and sources with high astrometric excess noise (HAEN). We select these stars by using Equation~\ref{eq:wd} that was originally used to filter them.

\begin{figure}
    \centering
    \includegraphics[width=\columnwidth]{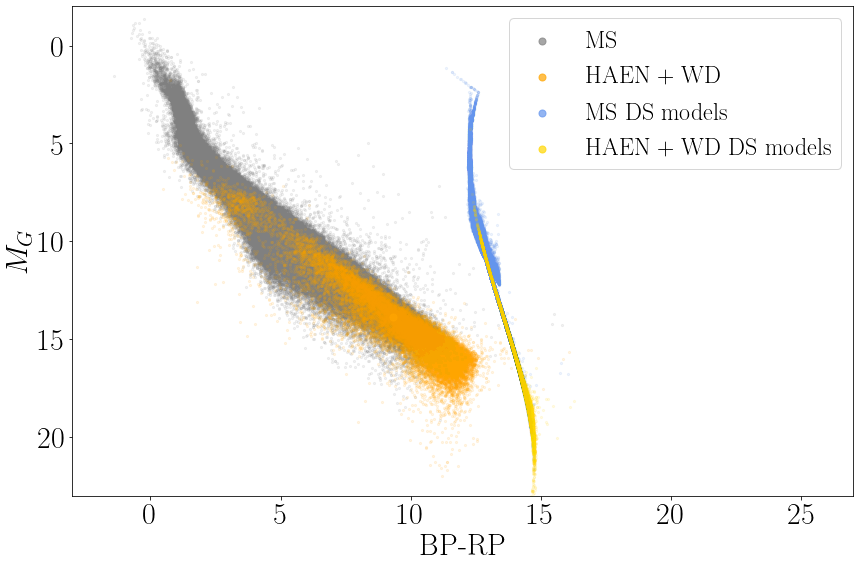}
    \caption{Dyson sphere models ($T_{\rm DS} = 100$ K, $\gamma = 0.9$) applied to main-sequence stars (MS), white dwarfs (WD), and sources with high astrometric excess noise (HAEN). Grey dots represent our main-sequence stars (MS), while blue dots are the models applied to these stars (MS DS). On the other hand, orange dots are white dwarfs and stars with high astrometric excess noise (HAEN + WD), while yellow dots are the models applied to these stars (HAEN + WD DS) }
    \label{fig:panels}
\end{figure}

We apply different DS models to all of these stars and compare their CMD signatures to models based on our sample of main-sequence stars. The result turns out to be quite similar, as illustrated in Figure~\ref{fig:panels} for white dwarfs and high astrometric excess noise stars, under the assumption of a $T_{\rm DS}$ = 100 K and $\gamma = 0.9$ DS model.

For red giant stars, the behavior is slightly different compared to the other sources. We define a star to be a red giant star if its Gaia magnitudes follow Equation~\ref{eq:postms}, originally used to filter them.

\begin{figure}
    \centering
    \includegraphics[width=\columnwidth]{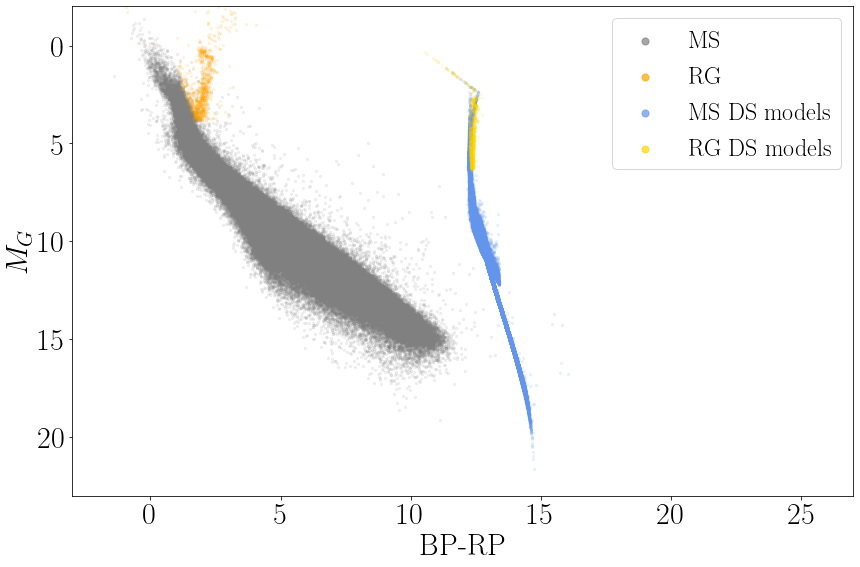}
    \caption{Dyson sphere models ($T_{\rm DS} = 100$ K, $\gamma = 0.9$) applied to main-sequence stars (MS) and red giant stars (RG). Grey dots represent our main-sequence stars (MS), while blue dots are the models applied to these stars (MS DS). On the other hand, orange dots are red giants (PMS), while yellow dots are the models applied to these stars (RG DS).}
    \label{fig:panels_pms}
\end{figure}

In Figure~\ref{fig:panels_pms}, we demonstrate how DS models based on red giant behave compared to those based on main-sequence stars, using the same DS parameters as in Figure~\ref{fig:panels}. As seen, the models for main-sequence stars and post-main-sequence stars partially agree. However, the boundary of the DS-compatible region changes in the regiome of very bright stars. Since we do not find a significant number of stars located in this region of any of the CMDs used, we conclude that the effect on our overall limits must me negligible.

\section{Luminosity fitting function}
\label{sec:lum}

The bolometric luminosity of a star is essential when modeling a system containing a DS. However, Gaia DR2 only provides luminosity estimates for a fraction of the of objects in the sample. To compensate for the lack of data, we model luminosities using a third-order polynomial fitting function. We choose main-sequence stars in the sample within 100 pc with luminosity estimations to generate this model. Since Gaia DR2 only provides luminosities for stars with $M_G$ < 10, we use PARSEC models \citep{bressan_parsec} to account for the missing data. PARSEC models correspond to theoretical stellar evolutionary tracks that provide stellar parameters as well as their photometry. We use the version 1.2S that incorporates the improvements presented in \citet{PC1,PC2,PC3}. From these models, we select isochrones of evolved systems (10$^{10}$ yr) to ensure the existence of low mass main-sequence stars. We also take tracks with metallicities between $Z = 0.012$ to $Z = 0.018$. This range should be representative of the thin-disc stars within 100 pc. We extend the range of stars with known luminosities by adding these models to $G$-band absolute magnitudes as faint as $\sim$ 15.5.

\begin{figure}
    \centering
    \includegraphics[width=\columnwidth]{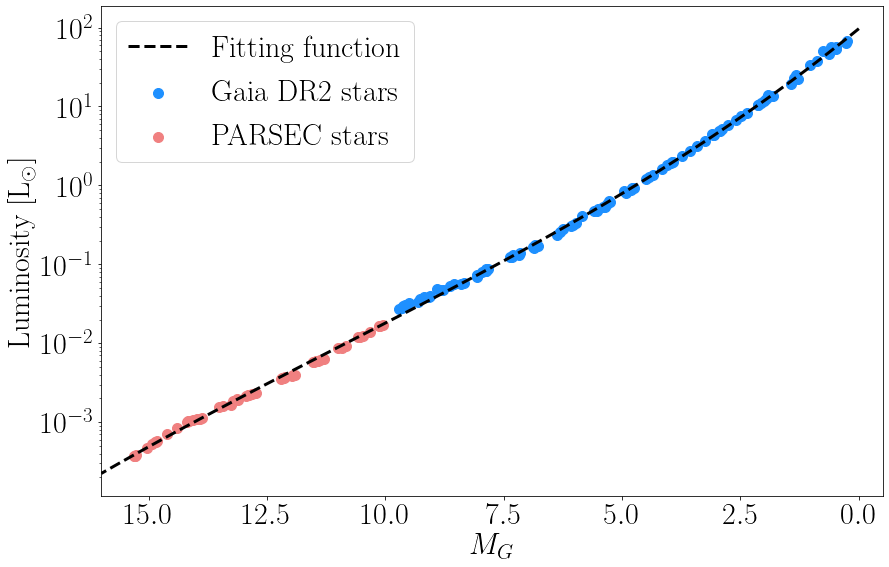}
    \caption{Luminosity as a function of the absolute magnitude in the $G$ band. The red dots correspond to the PARSEC model stars, while the blue dots correspond to the stars taken from the Gaia DR2 sample. The black line correspond to the fitting function.}
    \label{fig:fit}
\end{figure}

Using both types of luminosity estimates, we proceed to create a fitting function. However,  we count with more observational data estimates than theoretical estimations. To avoid a bias favouring the regime of massive stars, we base the fitting function on a pruned sample with equal number of stars in each $\Delta G$ = 0.5 bin. In generating this sample, we randomly stars from Gaia DR2 for which the luminosity uncertainties are lower than 1\%. We moreover do not consider stars brighter than $M_G = 0$ since their uncertainties tend to be higher. The resulting fitting function is described by  

\begin{equation}
\begin{aligned}
    \log(L/L_{\odot}) = & -4.98\times10^{-4} M_G^3 + 1.64\times10^{-2}M_G^2\\ 
    &-4.87\times10^{-1}M_G + 1.98,
\end{aligned}
\label{eq:fit}
\end{equation} where $L$ is the bolometric luminosity of the star and $M_G$ is its absolute magnitude in the $G$-band. The relation between the data used and the fitting function is shown in Figure~\ref{fig:fit}.

\section{Extinction effects}
\label{sec:extinction}
The DS models used in this paper ignore the effects of interstellar dust. In general, dust causes optical obscuration of sources yet excess emission at infrared wavelengths. Because of this, objects embedded in nebular regions tend to be common interlopers when checking the nature of sources compatible with DS models in the CMDs used throughout this paper.


Here, we explore the impact of dust obscuration on our models by comparing models with and without the extinction corrections provided for a subset of Gaia DR2 stars \citet{gaia_apsis}. 

We use a sample that contains data on stars within 5 kpc. In this sample, 83\% of them have $A_G$ values below 1, while 97\% of them have values below 2.

\begin{figure}
    \centering
    \includegraphics[width=\columnwidth]{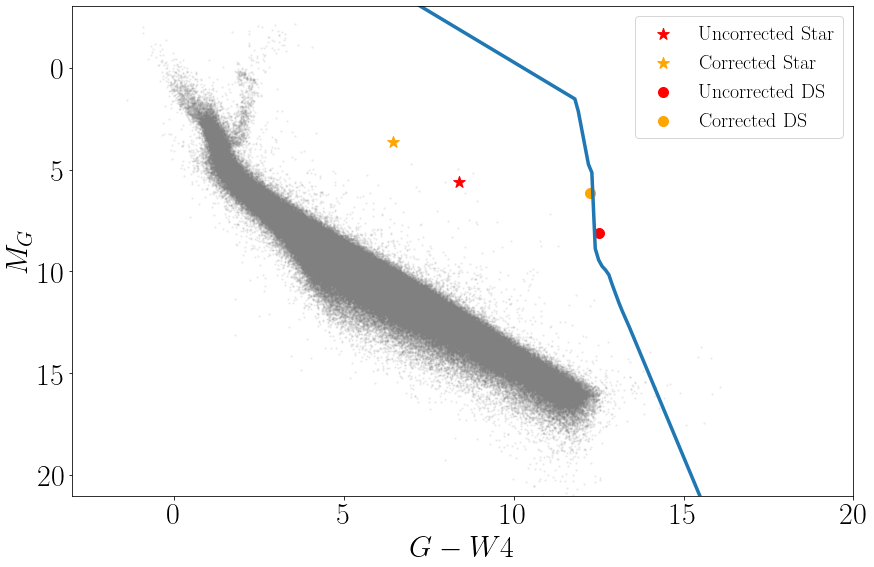}
    \caption{Colour-magnitude diagram showing the behaviour of DS models based on stars with and without extinction corrections. The red star represents a star with $A_G = 2.3$ prior to extinction correction and  the orange star the same object after correction. The red and orange dots represent DS models ($T_{\rm DS} = 300$, $\gamma = 0.9$) applied to these stars. Grey dots represent uncorrected stars within 100 pc. The blue line is the boundary that separates the DS and non-DS regions, derived from DS models based on stars not corrected for extinction.}
    \label{fig:ext}
\end{figure}

While these models differ due to the changes in the input luminosities and magnitudes of the stars, we find that the boundaries used to define DS-compatible regions within the CMDs use remain unchanged. In Figure~\ref{fig:ext}, we exemplify this for the star with the highest extinction estimate ($A_G = 2.3$ mag). As seen, the models with and without the extinctiion corrections both lie very close to the boundary. This trend holds for all the CMDs used in this paper.




\bsp	
\label{lastpage}
\end{document}